\documentclass[aps,prc,twocolumn,floatfix,nofootinbib,showpacs,superscriptaddress]{revtex4-1}
\usepackage[latin1]{inputenc}
\usepackage{bm}
\usepackage{epsfig}
\usepackage{amsthm}
\usepackage{amsfonts}
\usepackage{float}
\usepackage{amsmath,amssymb}
\usepackage{graphicx}% Include figure files
\usepackage{dcolumn}% Align table columns on decimal point
\usepackage{bm}% bold math
\def\nn{\nonumber}
\newcommand{\be}{\begin{equation}}
\newcommand{\ee}{\end{equation}}
\newcommand{\bea}{\begin{eqnarray}}
\newcommand{\eea}{\end{eqnarray}}

\newcommand{\om}{\omega}

\newcommand{\bk}{\boldsymbol{k}}
\newcommand{\bl}{\boldsymbol{l}}

\newcommand{\bpi}{\boldsymbol{\pi}}
\newcommand{\brho}{\boldsymbol{\rho}}

\newcommand{\del}{\partial}

\begin{document}
\title{Shear viscosity of pion gas due to $\rho\pi\pi$ and 
$\sigma\pi\pi$ interactions}
\author{Sabyasachi Ghosh}
\email{sabyaphy@ift.unesp.br}
\author{Gast\~ao Krein}
\email{gkrein@ift.unesp.br}
\affiliation{Instituto de F\'{\i}sica Te\'orica, Universidade Estadual Paulista,
Rua Dr. Bento Teobaldo Ferraz, 271 - Bloco II, 01140-070 S\~ao Paulo, SP, Brazil}
\author{Sourav Sarkar}
\email{sourav@vecc.gov.in}
\affiliation{Theoretical Physics Division, Variable Energy Cyclotron Centre, 
1/AF Bidhannagar, Kolkata 700064, India}
\begin{abstract}
We have evaluated the shear viscosity of pion gas 
taking into account its scattering with the low mass resonances,
$\sigma$ and $\rho$ during propagation in the medium.
The thermal width (or collisional rate) of the pions is calculated from
$\pi\sigma$ and $\pi\rho$ loop diagrams using effective
interactions in the real time formulation of finite temperature field theory.
A very small value of shear viscosity by entropy density ratio ($\eta/s$),
close to the KSS bound, is obtained which approximately matches the 
range of values of $\eta/s$ used by Niemi et al.~\cite{Niemi} in
order to fit the RHIC data of elliptic flow.
\end{abstract}
\maketitle
%
%
% 
%%%%%%%%%%%%%%%%%%%%%%%%%%%%%%%%%%%%%%%%%%%%%%%%%%%%%%%%%%%%%%%%%%%%%%%%%%%
\section{Introduction}
\label{sec:intro}

In order to explain the elliptic flow parameter, $v_2$, extracted from data
collected at the Relativistic Heavy Ion Collider (RHIC)~\cite{PHENIX1,STAR1,
PHENIX2,STAR2,BRAHMS,PHOBOS,PHENIX3}, hydrodynamical 
calculations~\cite{Romatschke1,Romatschke2,Heinz1,Heinz2,Roy} as well as some 
transport calculations~\cite{Xu1,Xu2,Greco1,Greco2} suggest that the matter 
produced in the collisions is likely to have a very small ratio of shear 
viscosity to entropy density, $\eta/s$.
Recent studies~\cite{Csernai,Gyulassy,Kapusta:2008vb,Chen1,Purnendu,Chen2} 
have shown that $\eta/s$ may reach a minimum in the vicinity of a phase 
transition - for earlier studies, see e.g.~Ref.~\cite{Hufner}. 
In this context, the smallness of this minimum value with respect to its lower 
bound, $\eta/s=1/4\pi$, commonly known as the KSS bound~\cite{KSS}, 
assumes particular significance. Again from the recent work of Niemi et 
al.~\cite{Niemi}, the transverse momentum $p_T$ dependence on elliptic 
flow parameter extracted from RHIC data is highly sensitive to 
the temperature dependence of $\eta/s$ in hadronic matter, and is almost 
independent of the viscosity in the QGP phase. This result attributes extra 
importance to the microscopic calculations of viscosity of hadronic matter in 
recent years~\cite{Dobado1,Dobado2,Muronga,Nakano,Nicola,Itakura,Gorenstein,Greiner,SPal,
Toneev,Prakash_2012,Buballa,Weise,SSS,Mitra,Prakash_2013}, though these 
investigations began some time ago~\cite{Gavin,Prakash}. 

Calculations based on kinetic theory (KT) approaches in Refs.~\cite{Gavin,Prakash,Toneev,SSS}
predict a shear viscosity $\eta$ of pionic matter that increases with~$T$, 
whereas using a Kubo approach, Lang et al.~\cite{Weise} predict $\eta$
to decrease with~$T$. For the interaction of pions in the medium, Lang 
et al.~\cite{Weise} used lowest order chiral perturbation theory ($\chi$PT), 
which describes well experimental data on $\pi-\pi$ cross sections up to
center-of-mass energies of $\sqrt s = 0.500$~GeV. For higher energies, resonances, 
particularly $\sigma$ and $\rho$, become important and iteration of the amplitude
(unitarization) is necessary to describe data. In the $\chi$PT approach, $\sigma$ 
and $\rho$ resonances in $\pi-\pi$ scattering can be generated dynamically under 
unitarization. Fernandez-Fraile et al.~\cite{Nicola} showed that under unitarization, 
$\chi$PT predicts $\eta$ increasing with $T$ in both Kubo and KT approaches -
without unitarization, $\eta$ decreases with $T$. Again in Ref.~\cite{Itakura}, it was 
shown that a KT approach leads to an $\eta$ of pionic medium that increases with $T$ 
when a phenomenological interaction used, while a decreasing function of $T$ is 
obtained when using $\chi$PT in that same approach. An increasing trend of $\eta$ with 
$T$ has also been observed by Mitra et. al.~\cite{SSS,Mitra}, who have incorporated a 
medium dependent $\pi-\pi$ cross-section in the transport equation for a pion gas. 
They also found a significant effect of a temperature dependent pionic chemical 
potential~\cite{Mitra}. Again, the question of magnitude of $\eta$ is also an unsettled 
issue. For example, near the critical temperature, $T_c \simeq 0.175$ GeV, 
Refs.~\cite{Itakura,Weise} predict an $\eta \approx 0.001$~GeV$^3$; in 
Refs.~\cite{Nicola,Prakash,SSS}, $\eta = 0.002-0.003$~GeV$^3$; and in 
Refs.~\cite{Dobado1,Dobado2}, $\eta = 0.4$~GeV$^3$. 

From these considerations, it is evident that the issue of the temperature dependence
of hadronic shear viscosity is still a matter of debate and warrants further
investigation. Motivated by this, we have calculated $\eta$ of a pion 
gas using an effective Lagrangian for $\pi\pi\sigma$ and $\pi\pi\rho$
interactions which may be treated as an alternative way to describe
$\pi -\pi$ cross sections up to the $\sqrt{s}=1$ GeV~\cite{SSS,Mitra}
beside unitarization technique~\cite{Nicola}. Using real-time thermal field theory 
we have calculated the in-medium pion correlator to obtain the
thermal width, a necessary ingredient to calculate
$\eta$. We have also estimated the temperature dependence of the shear viscosity 
to entropy density ratio $\eta/s$ of the pionic gas and compared our results
to others of the recent literature. Although the hadronic matter that is formed in
heavy-ion collisions at RHIC is comprised of more hadrons than pions only, our study 
nevertheless is of relevance to the real situation as, at least in the central rapidity 
region, pions are the dominant component of the hadronic fluid. 

In the next Section, we present the formalism used to evaluate the shear viscosity 
of a pion gas. Our numerical results are presented in Sec.~\ref{sec:num} and in 
Sec.~\ref{sec:concl}, we present the summary and conclusions.

%
%
% 
%%%%%%%%%%%%%%%%%%%%%%%%%%%%%%%%%%%%%%%%%%%%%%%%%%%%%%%%%%%%%%%%%%%%%%%%%%%
\section{Formalism}
\label{sec:form}

Let us start with the standard expression of the shear viscosity for pion gas: 
\be
\eta = \frac{\beta}{10\pi^2}\int\frac{d^3k\,\bk^6}{\Gamma_{\pi}(\bk,T)\,\om_k^2 } 
\, n(\om_k)\left[1+n(\om_k)\right] ,
\label{eta1_final}
\ee
where 
\be
n(\om_k) = \frac{1}{e^{\beta\om_k}-1} ,
\label{BEdist}
\ee 
is the Bose-Einstein distribution function for a temperature $T = 1/\beta$, with 
$\om_k = (\bk^2 +m_\pi^2)^{1/2}$, and $\Gamma_\pi(\bk,T)$ is the 
thermal width of $\pi$ mesons in hadronic matter at temperature~$T$. 
We note that this expression can be derived either with the Kubo formalism~\cite{Kubo}
using retarded correlator of the energy-momentum tensor, or with a kinetic approach
using the Boltzmann equation in the relaxation-time approximation~\cite{Reif}. In
both approaches, to evaluate $\Gamma_\pi(\bk,T)$ one needs the interactions of
the pions {\em in medium}. Here, we pursue the use of retarded correlators.

As mentioned previously, from the lowest order $\chi$PT, the estimated $\pi-\pi$ 
cross section in free space is well in agreement with the experimental data 
up to the center-of-mass energy $\sqrt{s}= 0.5$ GeV. Beyond this value of 
$\sqrt{s}$, the $\sigma$ and $\rho$ resonances play an essential role to 
explain the data. On unitarization, the $\sigma$ and $\rho$ resonances are 
generated dynamically~\cite{Nicola} in the amplitude. An alternative 
way, which we follow in the present paper, is to incorporate these resonances 
by using the effective interaction for $\pi\pi\sigma$ and $\pi\pi\rho$ 
interactions: 
\be
{\cal L} = g_\rho \, \brho_\mu \cdot \bpi \times \del^\mu \bpi 
+ \frac{g_\sigma}{2} m_\sigma \bpi\cdot\bpi\,\sigma,
\label{Lag}
\ee
where the coupling constants $g_\rho$ and $g_\sigma$ are fixed from
their experimental decay widths. We use this effective Lagrangian to calculate 
the contributions of the $\pi\rho$ and $\pi\sigma$ loops to the self-energy of 
$\pi$ meson at finite temperature. The contributions coming from the interactions 
of the pions in medium, which are the relevant ones for $\Gamma_\pi(\bk,T)$ 
in Eq.~(\ref{eta1_final}), can be obtained from the imaginary part of the retarded 
pion correlator $\Pi_{\pi}^R(k)$ evaluated at the $\pi$-meson pole, 
$k=(k_0=\omega_k,\bk)$. In~real-time thermal field theory, this relationship can 
be expressed as~\cite{Bellac,SG_NJL}:
\bea
\Gamma_\pi(\bk,T) &=& - \frac{1}{m_\pi} {\rm Im}\,{\Pi}_{\pi}^R(k)\vert_{k_0=\omega_k}
\nn\\[0.2true cm]
&=& - {\rm tanh}\left(\frac{\beta k_0}{2}\right) \frac{1}{m_\pi} 
{\rm Im}\,\Pi^{11}_{\pi}(k)\vert_{k_0=\omega_k}.
\label{R_bar_11}
\eea

\begin{figure}[h]
\begin{center}
\includegraphics[scale=0.6]{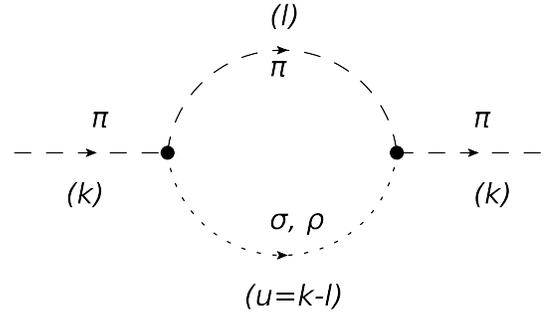}
\caption{One-loop self-energy diagram of pion.} 
\label{Leading_kind}
\end{center}
\end{figure}

For clarity of presentation, we start considering the correlator in the 
narrow-width approximation, in which the widths of the $\sigma$ and $\rho$ 
resonances are neglected. At one-loop order - see Fig.~\ref{Leading_kind} - 
one can write: 
\be
\Pi^{11}_{\pi}(k) = \Pi^{11}_{\pi}(k,\sigma) + \Pi^{11}_{\pi}(k,\rho) ,
\label{Pi11-pi}
\ee
with
\be
\Pi^{11}_{\pi}(k,u) = - i\int\frac{d^4l}{(2\pi)^4} \, L(k,l) \, 
D^{11}(l,m_l) \, D^{11}(u,m_u) ,
\label{Pi11}
\ee
for each loop ($\pi \sigma$ or $\pi \rho$), where $m_l = m_\pi$, $u=k-l$, 
and $m_u = m_\sigma$ for the $\pi\sigma$ loop and $m_u=m_\rho$ for the 
$\pi\rho$ loop; the propagators $D^{11}(l)$ are given by:
\be
D^{11}(l) = \frac{-1}{l^2-m^2_l + i\eta} + 2\pi i \,
n(\om_l)\, \delta(l^2-m^2_l) ,
\label{D11}
\ee
where $n(\om_l)$ is the Bose-Einstein distribution given in Eq.~(\ref{BEdist});
and
\be
L(k,l) = - \frac{g^2_\sigma m_\sigma^2}{4}, 
\ee
for the $\pi\sigma$ loop, and 
\bea
L(k,l) &=& -\frac{g^2_\rho}{m_\rho^2} \, 
\Bigl\{ k^2 \left(k^2 - m^2_\rho\right) + 
l^2 \left(l^2 - m^2_\rho\right)
\nn\\
&& - \, 2\left[ \left(k\cdot l\right) \, m^2_\rho + k^2 \,l^2 \right]
\Bigr\},
\eea
for the $\pi\rho$ loop. 

\vspace{1.0cm}

Using Eq.~(\ref{D11}) in Eq.~(\ref{Pi11}), one can perform
the $l_0$ integration and, from the relation between ${\rm Im}\,\Pi^R$ and
${\rm Im}\,\Pi^{11}$ in Eq.~(\ref{R_bar_11}), one obtains:

\begin{widetext}
\bea
{\rm Im}\,{\Pi}^{R}_{\pi}(k,u) &=& \int\frac{d^3l}{32\pi^2 \om_l\om_u}
\Bigl\{
L(k,l)\vert_{l_0 = \om_l}  \Bigl[\bigl(1 + n(\om_l) + n(\om_u)\bigr) 
\delta(k_0 -\om_l-\om_u)  - \,\bigl(n(\om_l) - n(\om_u)\bigr) 
\delta(k_0-\om_l+\om_u)\Bigr]
\nn\\[0.3true cm]
&& + \, L(k,l)\vert_{l_0=-\om_l} \Bigl[(n(\om_l)-n(\om_u)) \, \delta(k_0 +\om_l-\om_u) 
- \, \bigl(1 + n(\om_l) + n(\om_u)\bigr) \, \delta(k_0 +\om_l+\om_u)\Bigr]
\Bigr\}.
\label{self_LU}
\eea
\end{widetext}
The Dirac delta functions provide branch cuts in the $k_0$~-~axis, identifying 
the different kinematic regions where the imaginary part of the pion self-energy 
acquires non-zero values. The relevant term for the in-medium decay width
is the one proportional to $n(\om_l)-n(\om_u)$, which is due to the interactions 
of in-medium pions only and vanishes in vacuum. The relevant branch cut, the
Landau cut, is  the region  $- [\bk^2 + (m_u-m_\pi)^2]^{1/2} \leq k_0 
\leq [\bk^2 + (m_u-m_\pi)^2]^{1/2}$; it gives: 
\bea
\Gamma^{\rm nw}_\pi(\bk,T,u) &=& \frac{1}{16\pi|\bk| m_\pi} \int^{\om_-}_{\om_+} 
d\om \, L(\om) \nn \\[0.25true cm]
&& \times \, \left[n(\om) - n(\om_k + \om)\right],
\label{gm_int}
\eea
where the superscript ${\rm nw}$ indicates that this expression is obtained in the
narrow-width approximation, and
\be
\om_\pm = \frac{R^2}{2m_\pi^2} \left(- \om_k \pm |\bk| \, W \right) ,
\ee
with $R^2=2m_\pi^2-m_u^2$ and $W = \left(1- {4m_\pi^4}/{R^4}\right)^{1/2}$,
and
\be
L({\om}) = L(k_0=\om_k,\bk,l_0 = -\om,|\bl|=\sqrt{\om^2-m_\pi^2}).
\ee

The physical interpretation of the Landau cut contributions is 
straightforward~\cite{Weldon}. During propagation of $\pi^+$, it may 
disappear by absorbing a thermalized $\pi^-$ from the medium to create a 
thermalized $\rho^0$ or $\sigma$. Again the $\pi^+$ may appear by absorbing 
a thermalized $\rho^0$ or $\sigma$ from the medium as well as by emitting a 
thermalized $\pi^-$. $n_l(1+n_u)$ and $n_u(1+n_l)$ are the corresponding 
statistical probabilities of the forward and inverse scattering respectively.
By subtracting them, one gets the factor $(n_l-n_u)$
in Eq.~(\ref{gm_int}). 

Next, to take into account the widths of the resonances, we use the spectral
representations of the $\sigma$ and $\rho$ propagators in Eq.~(\ref{Pi11}) 
- see e.g. Refs.~\cite{S_rho,S_omega}. This results in a folding of the 
narrow-width expression for $\Gamma_{\pi}(\bk,T,m_u)$:
\be
\Gamma_{\pi}(\bk,T,m_u) = \frac{1}{N_u} \int^{(m^+_u)^2}_{(m^-_u)^2}
dM^2  \, \rho_u(M) \, \Gamma^{\rm nw}_\pi(\bk,T;M) ,
\label{gm_mu}
\ee
where $\Gamma^{\rm nw}(\bk,T;M)$ is the narrow-width expression given in Eq.~(\ref{gm_int}), 
with $m_u$ replaced by $M$; $\rho_u(M)$ is the spectral density:
\be
\rho_u(M) = \frac{1}{\pi}{\rm Im}\left[\frac{-1}{M^2-m_u^2+iM\Gamma_u(M)}\right],
\label{rho_u}
\ee
and $N_u$ is the normalization
\be
N_u = \int^{(m^+_u)^2}_{(m^-_u)^2}
dM^2\;\rho_u(M) .
\ee
$\Gamma_u(M)$, $u=\sigma,\rho$, are the spectral widths of the mesons:
\bea
\Gamma_\sigma (M) &=& \frac{3g_\sigma^2m_\sigma^2}{32\pi M} 
\left(1 - \frac{4m_\pi^2}{M^2}\right)^{1/2},
\label{s-width}
\\[0.25true cm]
\Gamma_\rho(M) &=& \frac{g_\rho^2M}{48\pi} 
\left(1 - \frac{4m_\pi^2}{M^2}\right)^{3/2}.
\label{r-width} 
\eea
In the integration limits, $m^\pm_u = m_u \pm 2\,\Gamma^0_u$, 
with $\Gamma^0_\sigma = \Gamma_\sigma(M=m_\sigma)$ and $\Gamma^0_\rho =\Gamma_\rho(M=m_\rho)$. 
In view of Eq.~(\ref{Pi11-pi}), the total pionic width is the sum
\be
\Gamma_\pi(\bk,T) = \Gamma_\pi(\bk,T,\rho) +  \Gamma_\pi(\bk,T,\sigma).
\label{totalGam}
\ee

A quantity closely related to the thermal width is the mean free path:
\be
\lambda_\pi(\bk,T) = \frac{|\bk|}{\om_k \, \Gamma_\pi(\bk,T)}.
\ee
Phenomenologically, analysis of this quantity is interesting for getting
further insight in the propagation of pions in medium; in particular, it
allows to know the values of typical pion momenta that are responsible 
for dissipation in medium, as we shall discuss in the next section. 
On~the theoretical side, this quantity is interesting as~\cite{Goity} 
$\lambda_\pi \equiv 1/\Gamma_\pi$ in the chiral limit, $m_\pi =0$; 
as such, the $m_\pi$ dependence of $\lambda_\pi$ provides insight on effects 
due to explicit chiral symmetry breaking~\cite{Weise}.
 
%
%
% 
%%%%%%%%%%%%%%%%%%%%%%%%%%%%%%%%%%%%%%%%%%%%%%%%%%%%%%%%%%%%%%%%%%%%%%%%%%%
\section{Results and Discussion}
\label{sec:num}

\begin{figure}[h]
\begin{center}
\includegraphics[scale=0.35]{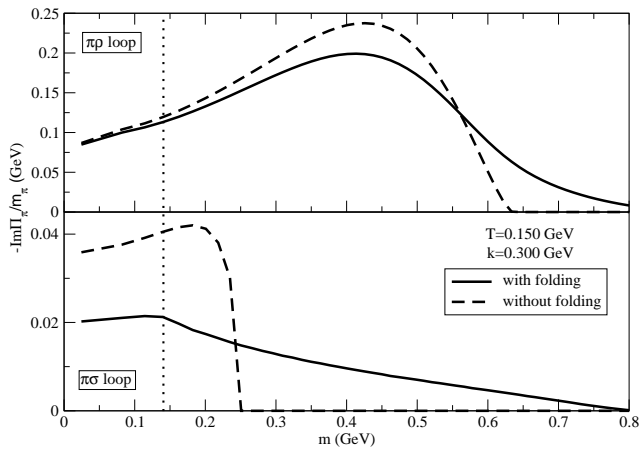}
\caption{The imaginary part of pion self-energy from $\pi\rho$ (upper panel) 
and $\pi\sigma$ (lower panel) loops as function of the invariant mass 
$m = \sqrt {k^2_0-|\bk|^2}$ for fixed values of temperature $T = 0.150$~GeV and 
three-momentum $| \bk | = 0.300$~GeV. The vertical dotted line indicates the 
on-shell value $m = m_\pi$. Parameters are: $m_\pi = 0.140$~GeV, 
$m_\rho = 0.770$~GeV, $\Gamma^0_\rho = 0.150$~GeV, $g_\rho = 6$ and Set~1 
in Table~\ref{tab} for parameters of $\sigma$ resonance. }
\label{fig:g_M}
\end{center}
\end{figure}

Let us first consider the separate contributions of the $\pi\rho$ 
and $\pi\sigma$ loops to the imaginary part of the pion self-energy as 
a function of the invariant mass $m^2=k_0^2-|\bk|^2$ for fixed values
of temperature, $T = 0.150$~GeV, and three-momentum, $|\bk| = 0.300$~GeV
- results are shown in Fig.~\ref{fig:g_M}. We have used here the following 
set of parameters: $m_\pi = 0.140$~GeV, $m_\rho = 0.770$~GeV, $\Gamma^0_\rho 
= 0.150$~GeV, and $g_\rho = 6$. The parameters for the $\sigma$ resonance 
are those of Set~1 in Table~\ref{tab}.

In Fig.~\ref{fig:g_M}, the dashed lines clearly 
indicate the sharp-ends of the Landau cuts at $m = m_\rho-m_\pi = 0.630$~GeV 
for the $\pi\rho$ loop (upper panel) and at $m = m_\sigma - m_\pi = 0.250$~GeV
for the $\pi\sigma$ loop (lower panel). These sharp ends turn into smooth
falloffs at large values of $m$ due to the folding with the spectral
functions of the $\sigma$ and $\rho$ resonances. This large-$m$ effect does 
not affect $\Gamma_\pi(\bk,T)$ as this quantity is calculated at $m = m_\pi$. 
However, folding does affect $\Gamma_\pi(\bk,T)$ via a large effect induced 
by the $\pi\sigma$ channel; at $m = m_\pi$,  folding decreases the contribution 
of the $\pi\sigma$ loop by $50$\% as compared to the corresponding contribution 
in the narrow-width approximation. This does not come as a surprise, as the
$\sigma$ resonance has a large width, while the width of the $\rho$ is not
as large. One should also notice that numerically, the contribution of the
$\rho$ resonance  to $\Gamma_\pi$ is one order of magnitude larger than the one 
from the $\sigma$ loop at $m = m_\pi$. However, as we shall see shortly, this 
does not mean that one can neglect the $\sigma$ resonance altogether.  

\vspace{0.5cm}

\begin{figure}[ht]
\begin{center}
\includegraphics[scale=0.35]{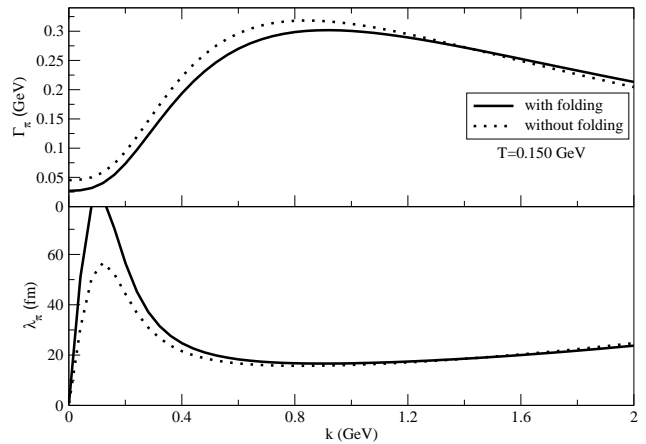}
\caption{Momentum dependence of the thermal width (upper panel) and of the 
mean free path (lower panel) for a fixed value of temperature, $T = 0.150$~GeV.
Parameters are the same as in Fig.~\ref{fig:g_M}.}
\label{fig:gm_pvec}
\end{center}
\end{figure}

Next, we consider the momentum dependence of thermal width and of the mean free 
path for a fixed temperature. Results are shown in Fig.~\ref{fig:gm_pvec}. First 
of all, one sees that the effects of folding are not big when considering the
joint contributions of the $\pi\rho$ and $\pi\sigma$ loops - this is due
to the combined facts that the width of $\rho$ has only a mild effect and the
dominance of the $\pi\rho$ loop over the $\pi\sigma$ loop. One also sees that 
the value of $\lambda_\pi$ is very big for momenta $0.100$ GeV $\le|\bk| \le 
0.300$~GeV, but for $|\bk| \ge 0.400$~GeV the value of mean free path varies
very little, reaching an average value of $\lambda_\pi \simeq 25$~fm. In a 
typical relativistic heavy ion collision at RHIC, the size of the hadronic
systems produced after freeze-out varies between $20$~fm and $40$~fm. Therefore, 
scattering processes with center of mass momenta larger than $|\bk| = 0.400$~GeV 
are those responsible for dissipation in the medium, at least for the chosen 
temperature $T = 0.15$~GeV. 
 
\vspace{0.5cm}

\begin{figure}[htb]
\begin{center}
\includegraphics[scale=0.35]{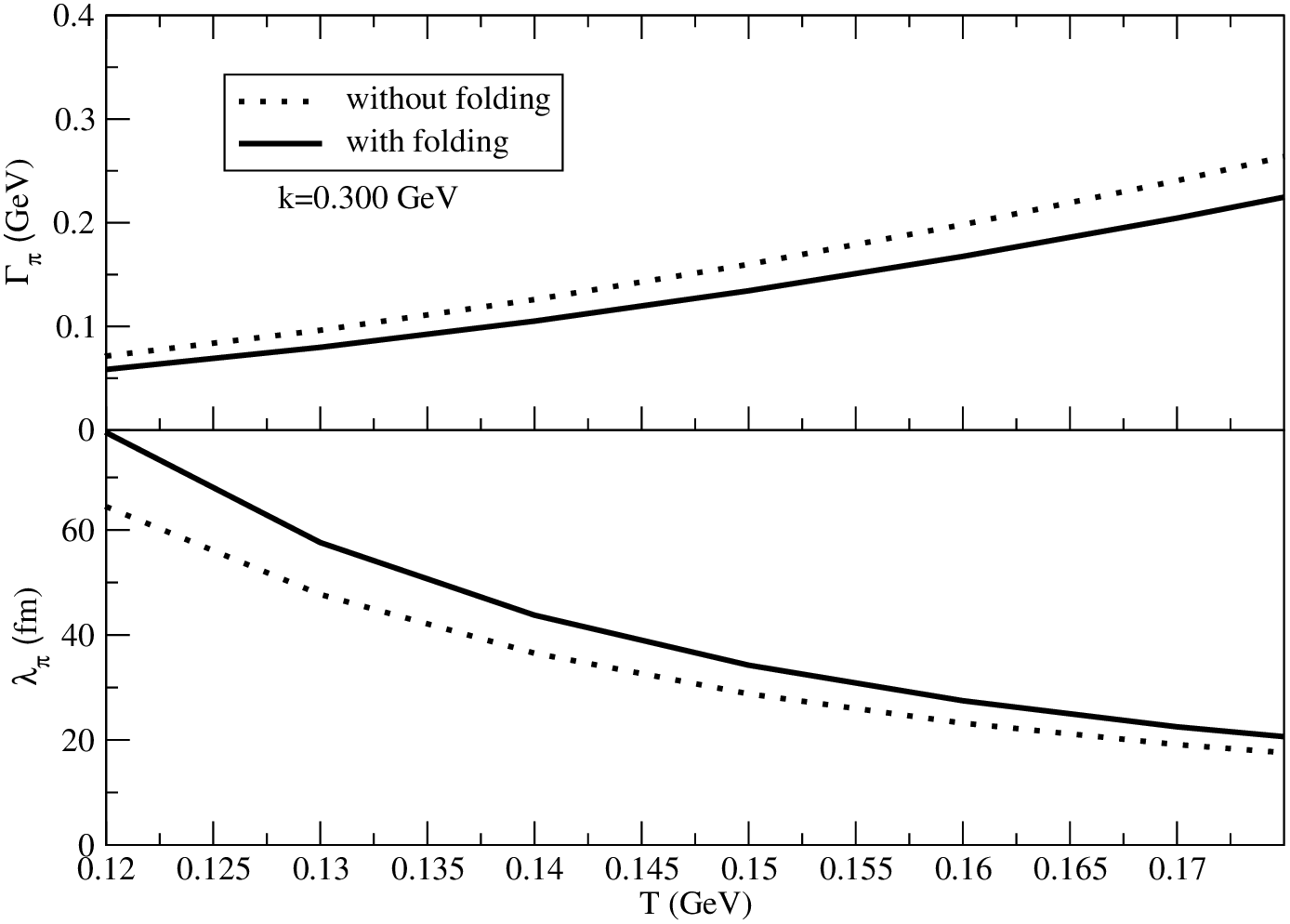}
\caption{Temperature dependence of the thermal width (upper panel) and of the 
mean free path (lower panel) for a fixed value of momentum $|\bk|=0.300$~GeV.
Parameters are the same as in Fig.~\ref{fig:g_M}.}
\label{fig:gm_T}
\end{center}
\end{figure}

In Fig.~\ref{fig:gm_T} we present results for the temperature dependence of the 
thermal width (upper panel) and of the mean free path (lower panel) for a fixed value 
of momentum $|\bk|=0.300$~GeV. Clearly, folding does not affect much the temperature 
dependence of these quantities; the reason for this is the same as for their 
momentum dependence: the dominance of the contribution of the $\pi\rho$ loop over 
that from $\pi\sigma$ loop. The figure also shows that only temperatures larger 
than $T = 0.120$~GeV give a mean free path smaller than the typical size of the 
hadronic system produced in a typical heavy ion collision at RHIC.

\vspace{0.5cm}

\begin{figure}[h]
\begin{center}
\includegraphics[scale=0.35]{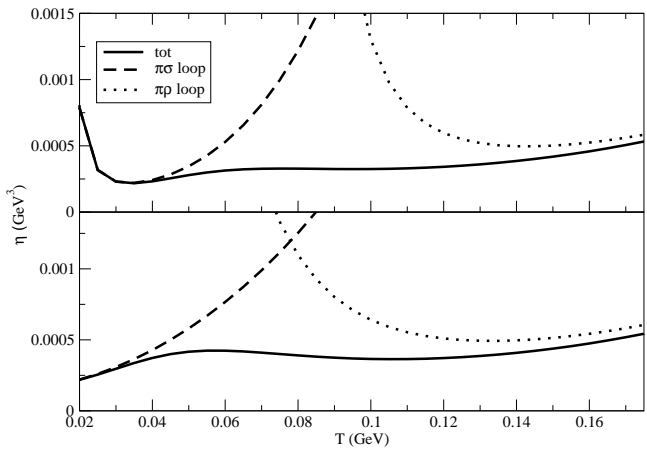}
\caption{Temperature dependence of $\eta$ from the $\pi\sigma$ (dashed lines)
and $\pi\rho$ (dotted lines) loops. The lower and upper panels respectively show 
the results with and without folding.}
\label{fig:eta_T_rs}
\end{center}
\end{figure}

Of course, the viscosity of the pion gas is determined not only by the value 
of $\Gamma_\pi$ (or $\lambda_\pi$), which is given basically by the 
$\pi-\pi$ interaction; it depends also on the momentum distribution 
of the in-medium pions, which is determined by the temperature in the 
Bose-Einstein distribution. In~Fig.~\ref{fig:eta_T_rs} we present the
results for the temperature dependence of $\eta$. Interestingly we see 
that the $\pi\rho$ and $\pi\sigma$ contributions play a complementary 
role in $\eta$ to be non-divergent in the higher ($T > 0.100$~GeV) and lower 
($T<0.100$ GeV) temperature regions respectively. The lesson here is that 
consideration of both resonances in $\pi-\pi$ scattering is strictly 
necessary to obtain a smooth, non divergent $\eta$ for temperatures below 
the critical temperature, $T_c \simeq 0.175$~GeV. Moreover, though $\eta$ at 
very low temperatures ($T<0.020$ GeV) tends to become very large in the narrow-width  
approximation (upper panel), this trend disappears after taking into account the
the widths of the resonances (lower panel).

\vspace{0.5cm}

We have compared our results with the earlier results in Kubo approach by 
Fernandez-Fraile et al.~\cite{Nicola} and Lang et al.~\cite{Weise}, along 
with previous results obtained by some of us~\cite{SSS} in a KT approach. 
In the KT approaches of Refs.~\cite{Prakash,Gavin,Toneev,SSS}, the predicted
$\eta$ is a monotonically increasing function of temperature in the temperature 
range 0.100 GeV $< T <$ 0.175 GeV and vanishing baryon chemical potential ($\mu=0$). 
The results of Lang et al.~\cite{Weise}
obtained with the Kubo approach indicate an $\eta$ decreasing in that same
temperature range. Similar trends are obtained by Fernadez-Fraile et al.~\cite{Nicola} 
with the Kubo-approach without unitarization of $\Gamma$, but the trend is reversed 
when dynamically generated (through unitarization) $\rho$ and $\sigma$ resonances 
come into play. Our calculations, based on an effective Lagrangian taking into 
account the low-mass $\sigma$ and $\rho$ resonances, found a similar trend of 
an increasing $\eta$ with $T$ for $T > 0.100$~GeV, although smaller in magnitude 
and slope, lending support to other calculations which take into account those
resonances.

\vspace{0.5cm}
\begin{figure}[htb]
\begin{center}
\includegraphics[scale=0.35]{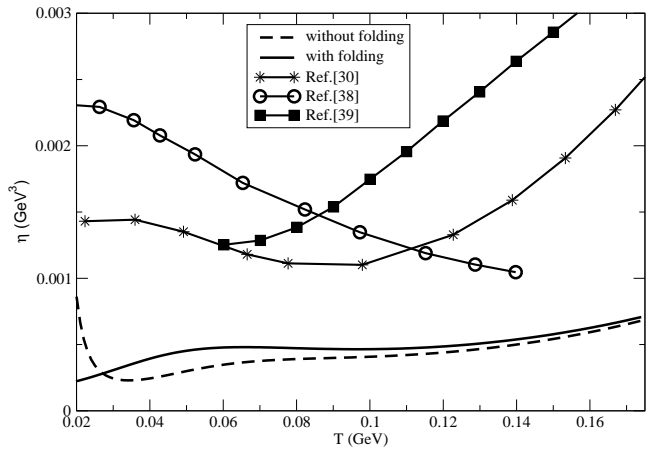}
\caption{Results of $\eta$ vs $T$ obtained in this work compared to some other results.}
\label{eta_T_all}
\end{center}
\end{figure}

\begin{figure}[htb]
\begin{center}
\includegraphics[scale=0.35]{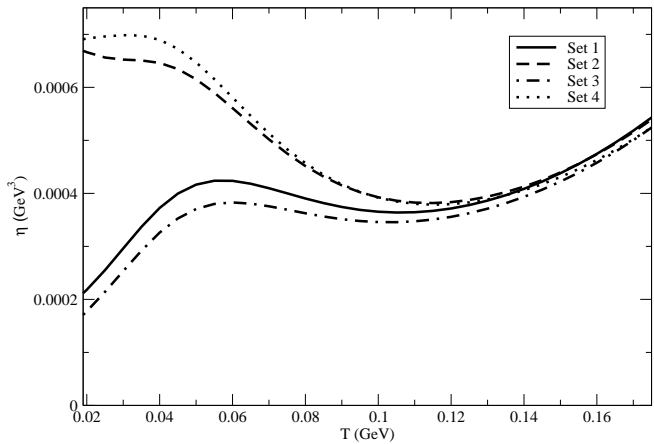}
\caption{The band of uncertainty of $\eta$ in the
low temperature domain for different sets of $m_\sigma$, $\Gamma(m_\sigma)$
and $g_\sigma$ from Table~\ref{tab}.}
\label{eta_T_band}
\end{center}
\end{figure}

Now we concentrate on the sensitivity of our predictions associated with phenomenological 
uncertainty of the parameters of the $\sigma$ resonance. The results presented above have
been obtained by choosing (arbitrarily) the parameters of Set~1 shown in 
Table~\ref{tab}. Although longstanding controversies about the properties of this 
resonance seem to be settling to a consensus~\cite{Pelaez}, recent literature~\cite{Mexico} 
still shows conflicting values for those properties, as one can see in Table~\ref{tab}.
We have explored the impact of the different values for the $\sigma$ parameters;
the results are shown in Fig.~\ref{eta_T_band}. As can be seen, all sets predict $\eta$
to be small, although parameter sets with smaller widths predict smaller $\eta$'s at 
low temperatures; for $T > 0.1$~GeV, all sets predict essentially the same result.

\begin{table}[htb]
\caption{The mass $m_\sigma$ (in GeV) and vacuum width $\Gamma^0_\sigma$ (in GeV) 
of the $\sigma$ resonance taken from Refs.~\cite{BES,E791,PDG}, from
which the corresponding coupling constants $g_\sigma$ are extracted.}
\begin{ruledtabular}
\begin{tabular}{l|ccc}
& & & \\ 
& $m_\sigma$ & $\Gamma^0_\sigma$ & $g_\sigma$ \\
& & &  \\
\hline
& & & \\
Set 1 (BES)~\cite{BES}     & 0.390    & 0.282    & 5.82  \\
& & & \\
Set 2 (E791)~\cite{E791}   & 0.489    & 0.338    & 5.73  \\
& & &  \\
Set 3 (PDG min)~\cite{PDG} & 0.400    & 0.400    & 6.85  \\
& & & \\
Set 4 (PDG max)~\cite{PDG} & 0.550  & 0.700  & 7.03  \\
& & & \\
\end{tabular}
\end{ruledtabular}
\label{tab}
\end{table}

\vspace{0.5cm}

\begin{figure}[htp]
\begin{center}
\includegraphics[scale=0.35]{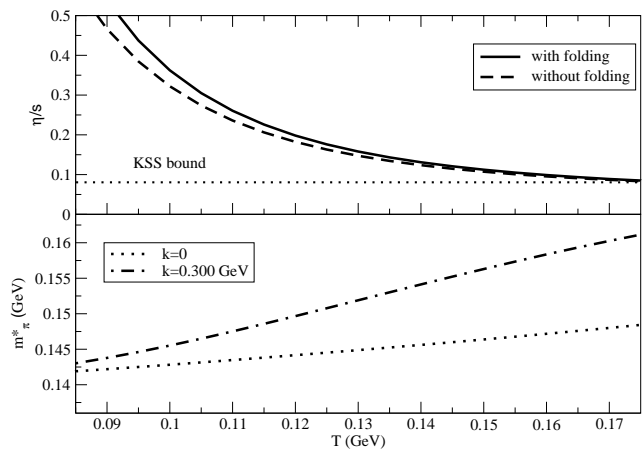}
\caption{Upper panel : $\eta/s$ vs $T$ and the KSS bound (dotted line).
Lower panel: Dependence of the effective pion mass $m^*_\pi$ with temperature 
$T$ at two different values of three momentum $|\bk|$.}
\label{eta_s}
\end{center}
\end{figure}

Finally, we estimate the temperature dependence of shear viscosity 
to entropy density ratio $\eta/s$ in
our model. In the calculation of the entropy density, 
\be
s = 3\beta\int \frac{d^3\bk}{(2\pi)^3} \left(\om_k+\frac{\bk^2}{3\om_k}\right)
n(\om_k),
\ee
where $\om_k=\sqrt{\bk^2 + m^{*2}_\pi(\bk,T)}$, we have explored the effect of 
the loops in the real part of $\Pi^R_{\pi}$ on the effective pion mass, 
$m^*_\pi=\sqrt{m_\pi^2+{\rm Re} \Pi^R_{\pi}(k_0=\om_k,\bk,T)}$. The variation 
of $m^*_\pi$ with $T$ for two different values of $\bk$ is shown in the lower 
panel of Fig.~(\ref{eta_s}). As can be seen, the effect is not big, at most 
15\% for the highest values of momentum and temperature. The effect of this 
change in the pion mass on the entropy density is marginal and can be safely
neglected.  

\vspace{0.5cm}

The dependence of the ratio $\eta/s$ on $T$ is shown in the upper panel of
Fig.~(\ref{eta_s}). Our results respect the KSS bound $\eta/s \le {1}/{4\pi}$,
as indicated by the dotted line.  We recall that Niemi et al.~\cite{Niemi} in their
investigation of $v_2(p_T)$ of RHIC data, have used an $\eta/s(T)$ from 
Ref.~\cite{Greiner} that is in the same range of our results shown in the figure.
This also lends support to the validity of the physical input our phenomenological 
analysis, in that the $\sigma$ and $\rho$ resonances play a decisive role in the 
dissipation properties of the pion gas.

%
%
% 
%%%%%%%%%%%%%%%%%%%%%%%%%%%%%%%%%%%%%%%%%%%%%%%%%%%%%%%%%%%%%%%%%%%%%%%%%%%
\section{Summary}
\label{sec:concl}

We have calculated the shear viscosity of a pion gas at finite temperature
taking into account the low mass resonances $\sigma$ and $\rho$ on pion
propagation in medium. The thermal width $\Gamma_\pi$ is calculated from 
one-loop pion self-energy at finite temperature in the framework of real-time 
thermal field theory. We have evaluated the contributions of $\pi\sigma$ and 
$\pi\rho$ loops to the pion self-energy with the help of an effective 
Lagrangian for the $\sigma\pi\pi$ and $\rho\pi\pi$ interactions. To take into
account the widths of $\sigma$ and $\rho$ resonances, we have folded the zero-widths
self-energies with their spectral functions. We have seen a complementary role 
played by the $\pi\sigma$ and $\pi\rho$ loops in producing a smooth temperature
dependence for $\eta$.  

We have also explored the impact of uncertainties in the parameters of the
$\sigma$ resonance on our results. Using the range of $\sigma$ mass 
($m_\sigma=0.400-0.550$ GeV) and width ($\Gamma_\sigma=0.400-0.700$ GeV) 
from the latest PDG compilation~\cite{PDG}, we have obtained smaller values 
for $\eta$ at low temperatures than those when using the earlier PDG 
values~\cite{PDG_old}, ($m_\sigma=0.400-1.200$ GeV and $\Gamma_\sigma=0.600-1.00$ 
GeV). For temperatures larger than $0.1$~GeV, all parameter sets give essentially
the same value for $\eta$

Our estimated temperature dependence for the ratio $\eta/s$ respects the KSS bound 
$\eta/s \le {1}/{4\pi}$, and comes very close to the bound for temperatures near the 
critical temperature $T_c = 175$~MeV. It agrees with the results of 
Refs.~\cite{Greiner,Gorenstein}. From the recent work by Niemi et al.~\cite{Niemi}, 
the elliptic flow parameter $v_2(P_T)$ of RHIC data prefers such small values of 
$\eta/s(T)$ for hadronic matter. The results seem to provide experimental justification 
to the microscopic calculations of shear viscosity which include $\sigma\pi\pi$ and 
$\rho\pi\pi$ interactions, as the one performed in the present work.

%
%
% 
%%%%%%%%%%%%%%%%%%%%%%%%%%%%%%%%%%%%%%%%%%%%%%%%%%%%%%%%%%%%%%%%%%%%%%%%%%%

\begin{acknowledgments}
Work partially financed by Funda\c{c}\~ao de Amparo \`a Pesquisa do Estado de 
S\~ao Paulo - FAPESP, Grant Nos. 2009/50180-0 (G.K.), 2012/16766-0 (S.G.), and
2013/01907-0 (G.K.); Conselho Nacional de Desenvolvimento Cient\'{\i}fico e 
Tecnol\'ogico - CNPq, Grant No. 305894/2009-9 (G.K.). 
\end{acknowledgments}

%
%
% 
%%%%%%%%%%%%%%%%%%%%%%%%%%%%%%%%%%%%%%%%%%%%%%%%%%%%%%%%%%%%%%%%%%%%%%%%%%%

\end{document}